\documentclass[twocolumn]{aastex61}

\shorttitle{Flare Modulation of Three- and Five-Minute Oscillations in Studying Wave Propagation}
\shortauthors{A.~Chelpanov, N.~Kobanov}
\begin{document}
\title{Using Flare-Induced Modulation of Three- and Five-Minute Oscillations for Studying Wave Propagation in the Solar Atmosphere}
\correspondingauthor{A.~\surname{Chelpanov}}
\email{chelpanov@iszf.irk.ru}
\author{Andrei~\surname{Chelpanov}}
\affil{Institute of Solar-Terrestrial Physics
                     of Siberian Branch of Russian Academy of Sciences, Irkutsk, Russia}
\author{Nikolai~\surname{Kobanov}}
\affiliation{Institute of Solar-Terrestrial Physics
                     of Siberian Branch of Russian Academy of Sciences, Irkutsk, Russia}

\begin{abstract}

We propose a method for diagnosing the physical conditions in the solar atmosphere using an increase in oscillation amplitudes resulting from minuscule solar flares.
As an example, we consider a B2 flare, which caused a sharp short-lived increase in the amplitude of three- and five-minute oscillations in the lower layers of the solar atmosphere.
Enhanced three- and five-minute oscillations propagated from the lower layers of the atmosphere into the corona.
Such short oscillation trains made it possible to remove the uncertainties arising in the measurements of the phase and group lags between the layers.
In addition, the amplification of the oscillations that reach the corona may add to the likelihood of a repeated flare.
Studying oscillations in small flare events has the advantage of exploring the atmosphere in its quasi-quiet condition as opposed to powerful flares, which cause substantial and prolonged disturbance of the environment.
In addition, small flares are much more common than powerful flares, which allows one to choose from a larger sample of observational material.
 
\end{abstract}

\section{Introduction} \label{sec:intro}

Solar flares often cause oscillatory and wave processes in the solar atmosphere.
Quasiperiodic pulsations (QPP), the most known type of such processes,  are occasionally observed in the extreme ultraviolet with quasi-periods from seconds to tens of seconds \citep{2009SSRv..149..119N, 2016SoPh..291.3143V, 2021SSRv..217...66Z}.
Originating in the upper layers of the solar atmosphere, these oscillations propagate downwards, even reaching the chromosphere \citep{2018SoPh..293..157C, 2020ARep...64..363C}.
The impact of flares on oscillations, however is not limited to the generation of QPPs.
Flare-induced perturbations of the lower solar atmosphere either make the environment oscillate at its natural frequencies, or sharply increase the power of the oscillations already existing in it \citep{2016AJ....152...81M, 2017ApJ...848L...8M, 2019SoPh..294...58K, 2020ApJ...903...19F}.
Such oscillations in the solar photosphere and chromosphere are represented by five- and three-minute periods, respectively.
A several-fold increase in the amplitude of these oscillations following a flare is usually clearly localized in time and space \citep{2018SoPh..293..157C, 2020ApJ...903...19F}.
This confirms the causal relationship of this phenomenon with solar flares.

In the studies of the physical characteristics in different layers of the solar atmosphere, measurements of the time lag between the oscillations at selected frequencies is widely used \citep{1978SoPh...58..347G,1984ApJ...277..874L, 2004AstL...30..489G,2006ASPC..358..465C,2009ApJ...692.1211C, 2010ApJ...718...11G, 2013SoPh..284..379K,2016SoPh..291.3349T,2018SoPh..293....2D}. Measuring the time lags between signals from different heights of the atmosphere is the main means to study wave propagation processes \citep{1983A&A...123..263U,1989A&A...213..423D,2006ASPC..358..465C,2013SoPh..284..379K,2015ARep...59..959D}.
This information may help to calculate wave-propagation velocities or determine more accurately the formation height of the used spectral lines.
Therefore, any progress in this direction aids better understanding of the energy-exchange processes between the layers of the solar atmosphere.

In this work, we propose to use flare-induced modulation of three- and five-minute acoustic oscillations in order to measure the time of their propagation to the upper layers of the solar atmosphere.

\section{Instruments and Data}

In the analysis, we used spectral observations at the location of a small flare that occurred on 21\,September\,2012 in a facula in AR\,115736 close to disk centre (SOL2012-09-21T02:19).
The magnetic structure of the facula consisted of a dipole, and the flare was located close to the polarity inversion line.
Based on the X-ray flux from the \textit{Geostationary Operational Environmental Satellite} \citep{GOES, 1996SPIE.2812..344H}, we determined the flare to start, peak, and end at 02:12\,UT, 02:19\,UT, and 02:24\,UT, although the chromospheric lines exhibited effects of the flare up until 02:30\,UT.
\citet{2018JASTP.173...50K} study the flare in more detail.

At the time of the flare, we were observing the active region at the ground-based \textit{Horizontal Solar Telescope} of the \textit{Sayan Solar Observatory}.
The observations included spectrograms on the facula in the photospheric Si\,\textsc{i} 10827\,\AA\ line and two chromospheric lines: H$\alpha$ 6563\,\AA\ and He\,\textsc{i} 10830\,\AA.
The spectrograms’ cadence is 1.5\,seconds, and the spectral resolution is 5\,--\,16\,m\AA\ depending on the wavelength.
The real spatial resolution of the observations is limited by the atmosphere’s disturbances; we estimated it to be 1.0\,--\,1.5\,arcsec.
From the spectrograms, we derive intensity and line-of-sight velocity signals.
The length of the series is 100\,minutes, and the flare occurred approximately in the middle of the series.

The \textit{Atmospheric Imaging Assembly} onboard the \textit{Solar Dynamics Observatory} (SDO/AIA) 304\,\AA\ and 171\,\AA\ channels provide emission-intensity observations in the transition region and lower corona, respectively.
These data have a cadence of 12\,seconds; the spatial resolution is 0.6 arcsec per pixel.
For the analysis, we used the data from the same period that is covered in our ground-based observations.

To extract the intensity and velocity signals from the spectral data, we normalized the line profiles to the continuum intensity in order to compensate for variations of non-solar origin.
The velocity signals were obtained with the use of the lambda-meter method \citep{1967AnAp...30..257R}.
This method suggests measuring the line displacements by determining the position of two segments of the opposite line wings with equal intensity and a given distance between them. This distance is constant for each frame throughout the series. The same line profile segments were used to obtain the intensity signals.

The SDO data were prepared and de-rotated with the use of the algorithms available in the \textit{SunPy core package} \citep{sunpy_community2020}.
The SDO light curves were chosen from the locations corresponding to the position of the southern part of the slit where the flare occurred.

For the analysis, we used narrow-band frequency filtration based on a Morlet wavelet of the sixth order \citep{1998BAMS...79...61T}. The typical band width that we used was 1\,mHz.

\section{Results and Discussion}

\begin{figure}
\centerline{
\includegraphics[width=8cm]{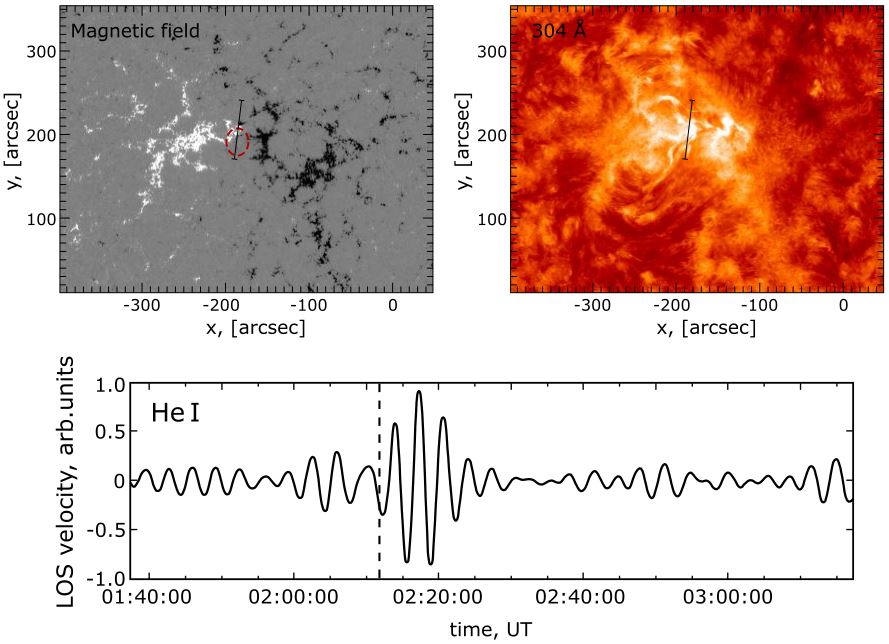}
}
\caption{\textit{Upper panels:} the location of the spectrograph slit and the flare site (dashed circle) shown against the SDO/HMI magnetogram of the AR and the SDO/AIA 304\,\AA\ channel image at 02:19\,UT; \textit{Bottom panel:} oscillations of the chromospheric LOS velocity signal filtered in the 5\,--\,6\,mHz frequency range before, during, and after the flare. The \textit{vertical dashed line} marks the beginning of the flare.}
\label{fig:0}
\end{figure}

In this work, we use observations of a facula, during which a small flare occurred in the spectrograph aperture (Figure~\ref{fig:0}).

\begin{figure}
\centerline{
\includegraphics[width=9cm]{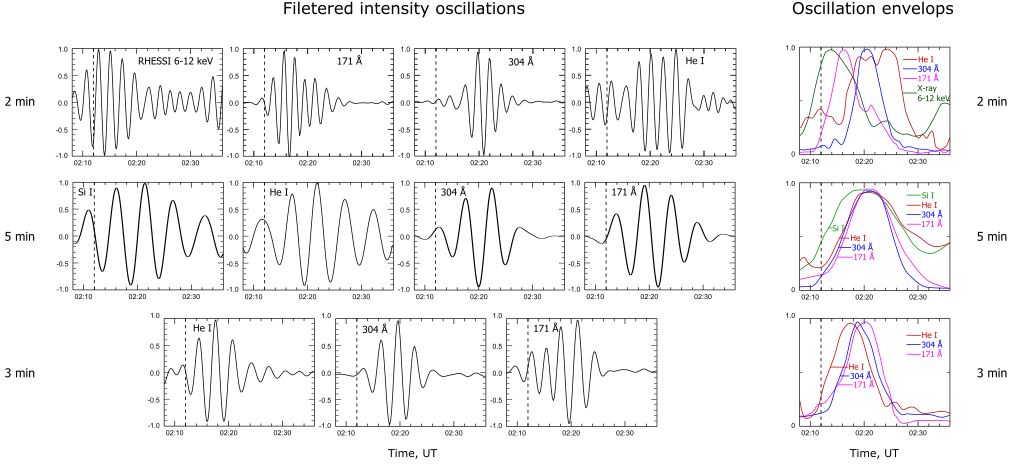}
}
\caption{\textit{Left:} oscillation trains of different periods; \textit{Right:} oscillation train envelopes.}
\label{fig:1}
\end{figure}

Oscillations of all the observed layers of the solar atmosphere (photosphere, chromosphere, transition zone, and corona) reacted to the flare with a short-term increase in the oscillation amplitude (Figure~\ref{fig:0}, bottom panel).
To isolate oscillations with a given period from the signal, we applied narrow-range wavelet filtering.
The right-hand column in Figure~\ref{fig:1} shows the envelopes of the oscillation wave trains.
The oscillation trains lasted for 15\,--\,20 minutes for different periods and different heights in the atmosphere.

The first oscillatory response to the flare manifested itself in two-minute oscillations.
The beginning of the oscillatory train successively appeared in X-rays, then in the 171\,\AA\ and 304\,\AA\ channels, and in the He\,\textsc{i} line.
Thus, the two-minute oscillation train appears later and later with altitude decreasing from the reconnection site to the chromosphere.
Hence, there is a conclusion that these oscillations propagated downwards.

In the case of oscillations with longer periods -- three and five minutes, which are more typical of the lower atmosphere -- the opposite picture is observed: the higher a layer is formed, the later the oscillations appear in it, i.e. they propagate upwards.
For five-minute oscillations, the delay is less pronounced. This is probably due to the fact that five-minute oscillations always dominate in the photosphere and chromosphere of faculae \citep{1965ApJ...141.1131O, 1971SoPh...19..338S, 2008ApJ...676L..85K, 2011SoPh..268..329K, 2017A&A...598L...2K}, while three-minute oscillations are not so typical of them.
We can assume that, along with the propagating ones, the standing waves make a large contribution to the five-minute signal \citep{1974SoPh...39...31D, 1999SoPh..184..253G, 2011SoPh..273...39K}.
One more reason for the less consistent five-minute oscillation propagation pattern is that our measurements are made for waves propagating in a strictly vertical direction, while five-minute oscillations are more likely to propagate along inclined magnetic loops.

Thus, we can conclude that the flare modulated the oscillation signals in the underlying atmosphere \citep{2018SoPh..293..157C}.
When the disturbance from the flare reaches the lower layers, they continue to oscillate at their typical natural frequencies, and the disturbance serves as an impetus that amplifies these natural oscillations.

Note that the wave propagation from the lower layers upwards is usually not registered when studying flare phenomena.
In powerful flares -- and such flares are examined in the overwhelming majority of works devoted to flares -- magnetic reconnections occur in the upper layers of the solar atmosphere, and after that the effect from flares spreads to the lower layers in the form of radiation, MHD waves, and particle fluxes.
Therefore, practically all works devoted to wave propagation in flare events study the downward propagation \citep{2008ApJ...675.1645F, 2016SoPh..291...89V,2018ApJ...867...85B}.

Two-minute oscillations in the upper layers are amplified immediately with the onset of the flare.
Three- and five-minute oscillations exist in the lower facular atmosphere under normal undisturbed conditions, and the formation of the amplified oscillation trains there starts noticeably later after the onset of the flare, i.e. when the effect of the flare reaches the lower layers.

\begin{figure}
\centerline{
\includegraphics[width=7cm]{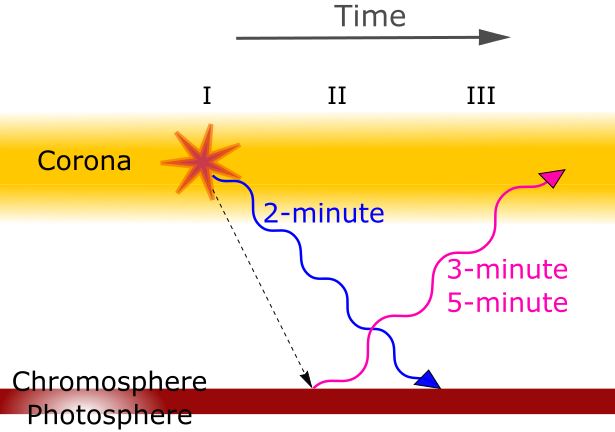}
}
\caption{Schematic representation of the sequence of events. The arrow marks the passage of time from left to right: the reconnection in the corona~(I) produces the two-minute oscillation train, which propagates downwards along with the agent (\textit{dotted line}) that later causes an increase in three- and five-minute oscillations (II); the three- and five-minute oscillation trains propagate upwards to the corona (III).}
\label{fig:2}
\end{figure}

Identifying the agent that directly caused the amplification of the three- and five-minute oscillations in the lower layers, we can exclude the two-minute oscillations propagating downwards.
This follows from the fact that a sharp increase in the three- and five-minute oscillations begins noticeably earlier than the two-minute oscillations reach the chromosphere (Figure~\ref{fig:2}).
We may assume that such an agent is a stream of accelerated particles.
\citet{2020ApJ...903...19F} came to the same conclusion based on the analysis of the relation between three-minute oscillations in the sunspot active region after a flare.
During the flare, we observed a dimming in the He\,\textsc{i}\,10830\,\AA\ line, which may have as well resulted from energetic particles \citep{2021ApJ...912..153K}.

Oscillation trains caused by a small flare propagating through the atmospheric layers provide a convenient opportunity for diagnosing the physical conditions in the atmosphere.
Such modulation makes it possible not only to estimate the group velocity of waves based on the envelope delays, but also to overcome the difficulties associated with the uncertainty of the phase shift between the signals.
This is because, first, in such cases the wave train is well pronounced due to the significant amplification over the background (see Figure~\ref{fig:0}), and, second, it is short-lived with its beginning and end easily traced, and during this time the phase difference between the signals does not have time to change significantly.
 
\begin{figure}
\centerline{
\includegraphics[width=5cm]{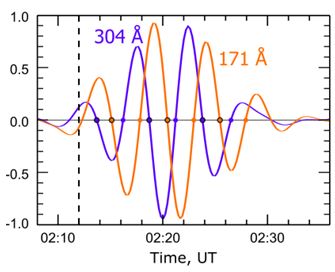}
}
\caption{The phase delay is determined based on the points where the signal crosses the zero line. The time-shift value of the wave trains is calculated as the average value over the time intervals between these crossings.}
\label{fig:3}
\end{figure}

To assess the phase shift between the signals, it is proposed to specify the moments when the signal crosses the zero line (Figure~\ref{fig:3}).
The time-shift value is calculated as the average over all pairs of adjacent points belonging to different signals.

In the case of this flare, the average phase delay for signal pairs Si\,\textsc{i}\,--\,He\,\textsc{i}, He\,\textsc{i}\,--\,304\,\AA, and 304\,\AA\,--\,171\,\AA\ were 49$\pm$25\,seconds, 38$\pm$16\,seconds, and 103$\pm$6\,seconds.

A significant advantage of using small flares, as opposed to large flares, is that it allows one to study the behavior of waves under conditions close to normal conditions existing in the quiet Sun, while the energy released in large flares disturbs the environment for a long time.
In addition, small flares offer more opportunities, since they occur more often than larger flares.

It should be noted that a significant increase in the amplitude of the oscillations propagating upwards increases the likelihood of a magnetic reconnection, leading to a repeated flare.
In this regard, the question arises: what role do small, repeated flares play? Are they analogous to a safety valve for dumping excess energy, or, on the contrary, do they contribute to the development of a larger outbreak? This is a subject for future research.

\section{Conclusion}

We propose to use the flare-caused modulation of the natural oscillations in the lower solar atmosphere when studying the processes of wave propagation upwards into the transition zone and corona.
In particular, this allows avoiding uncertainties when measuring phase shifts between signals at different altitudes.
One of such uncertainties comes from the difficulty in assessing the number of periods included in the phase shift: in the cases when the wave trains are not well pronounced it is often impossible to unambiguously determine whether the phase shift is greater than one oscillation period or not.
Often, wave trains propagating between layers do not show the same shape at different heights, which further hinders the understanding of how to evaluate the propagation speed.
A small flare induces a well-defined wave train, whose beginning, peak, and end are easily traced in all the studied heights.
Such trains are short-lived, which helps reduce the errors associated with the change of the phase during the train.
This helps to clearly determine the phase shifts between the height levels.
One can see some analogy with geophysics, when small blasts in boreholes are used to study the properties of the underlying rocks.
At the same time, the explosive disturbance also sharply activates oscillations at natural frequencies in the rocks surrounding the well. We believe that the proposed method will be useful for future research, since minuscule flares occur much more often than medium- and high-power flares \citep{2021ApJS..253...29L}, hence the abundance of available observational data.

We determined the time lags for the line pairs Si\,\textsc{i}\,--\,He\,\textsc{i}, He\,\textsc{i}\,--\,304\,\AA, and 304\,\AA\,--\,171\,\AA\ to be 49$\pm$25\,seconds, 38$\pm$16\,seconds, and 103$\pm$6\,seconds. Our results approximately correspond to those found earlier \citep{2012A&A...543A...9Y,2013SoPh..284..379K}, although they contradict some other findings. For example, \citet{2006ASPC..358..465C} measured the phase shifts between the Si\,\textsc{i} and He\,\textsc{i} lines to be 390\,seconds and 330\,seconds for five- and three-minute oscillations in a facula.

Another aspect of this phenomenon is that flare stimulation of the oscillation power propagating upwards increases the likelihood of magnetic-loop reconnection and may contribute to the occurrence of a new flare.
In this regard, detailed characteristics of oscillations in repeated flares are of interest for future studies.

\acknowledgments

The reported study was funded by RFBR, project number 20-32-70076 and Project No.\,II.16.3.2 of ISTP SB RAS. Spectral  data were recorded at the Angara Multiaccess Center facilities at ISTP SB RAS. We acknowledge the NASA/SDO and RHESSI science teams for providing the data.
We thank the anonymous reviewer for the helpful remarks.

\bibliography{Chelpanov}

\end{document}